\AtBeginDocument{%
  \paperwidth=\dimexpr
    1in + \oddsidemargin
    + \textwidth
    + 1in + \oddsidemargin
  \relax
  \paperheight=\dimexpr
    1in + \topmargin
    + \headheight + \headsep
    + \textheight
    + 1in + \topmargin
  \relax
  \usepackage[pass]{geometry}\relax
}
\RequirePackage{fix-cm}

\documentclass{svjour3}
\smartqed  

\usepackage{graphicx}
\usepackage{pifont}
\usepackage{fancybox}
\usepackage[hidelinks]{hyperref}
\usepackage{url}
\usepackage{multirow}
\usepackage{tablefootnote}
\usepackage{caption}
\usepackage{makecell}
\usepackage{balance}
\usepackage{upquote}
\usepackage[utf8]{inputenc}
\usepackage[usenames,dvipsnames,svgnames,table]{xcolor}
\usepackage{color}
\usepackage{fancyref}
\usepackage{amsmath}
\usepackage{amssymb}
\usepackage{threeparttable}

\let\oldding\ding

\renewcommand{\ding}[2][1]{\scalebox{#1}{\oldding{#2}}}

\newcolumntype{P}[1]{>{\centering\arraybackslash}p{#1}}

\usepackage{subfig}
\usepackage[pscoord]{eso-pic}
\newcommand{\placetextbox}[3]{
  \setbox0=\hbox{#3}
  \AddToShipoutPictureFG*{
    \put(\LenToUnit{#1\paperwidth},\LenToUnit{#2\paperheight}){\vtop{{\null}\makebox[0pt][c]{#3}}}%
  }%
}%

\journalname{The Journal of Supercomputing}

\begin{document}

\title{Loginson: a transform and load system for very large scale log analysis in large IT infrastructures}
\placetextbox{0.76}{0.77}{\parbox{0.55\textwidth}{ \footnotesize{
    This is a version of an unedited manuscript that accepted for publication. Please, cite as: \\ \\
    \textbf{Vega, C., Roquero, P., Leira, R. et al., Loginson: a transform and load system for very large-scale log analysis in large IT infrastructures, The Journal of Supercomputing (2017)} \\ \\ The final publication is available at Springer via: \url{http://dx.doi.org/10.1007/s11227-017-1990-1}
}}}
\titlerunning{\includegraphics[height=0.5cm]{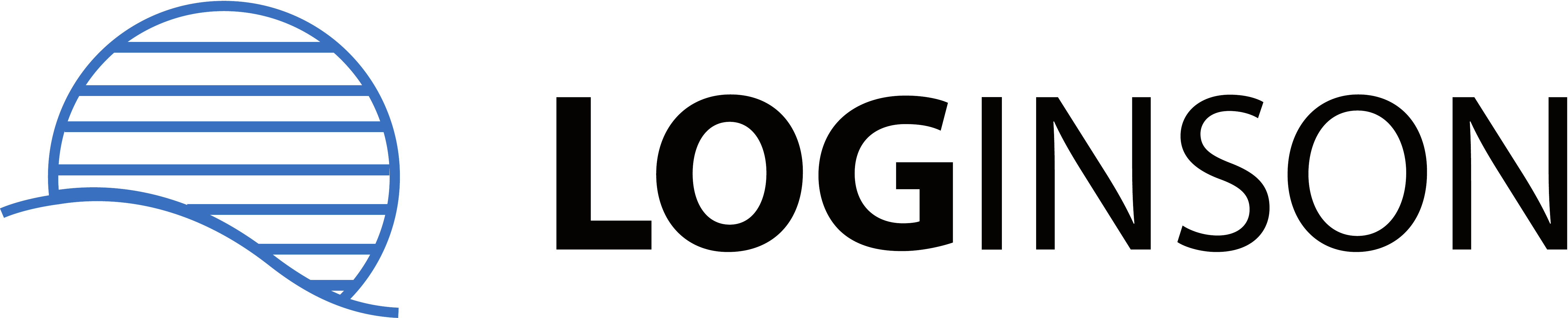}}        
\author{Carlos Vega \and Paula Roquero \and Rafael Leira \and Ivan Gonzalez \and Javier Aracil }
\institute{Departamento de Tecnolog\'{i}a Electr\'{o}nica y de las Comunicaciones, E.P.S.\\
Universidad Aut\'{o}noma de Madrid. 28049 Madrid, Spain\\
\email{carlosgonzalo.vega@predoc.uam.es}\\Tel.: (+34) 91 497 2839}

\date{Received: 27 Jun 2016 / Accepted: 22 Feb 2017}

\maketitle

\begin{abstract}
\label{abstract}
Nowadays most systems and applications produce log records that are useful for security and monitoring purposes such as debugging programming errors, checking system status and detecting configuration problems or even attacks. To this end, a log repository becomes necessary whereby logs can be accessed and visualized in a timely manner.

This paper presents {\em Loginson}, a high performance log centralization system for large-scale log collection and processing in large IT infrastructures. Besides log collection, {\em Loginson} provides high level analytics through a visual interface for the purpose of troubleshooting critical incidents.

We note that {\em Loginson} outperforms all of the other log centralization solutions by taking full advantage of the vertical scalability, and therefore decreasing Capital Expenditure (CAPEX) and Operating Expense (OPEX) costs for deployment scenarios with a huge volume of log data.
\keywords{Log analysis \and operational intelligence \and data repositories for log collection \and large datacenters}
\end{abstract}

\section{Introduction}
\label{sec:intro}

In large datacenters and, in general, IT infrastructures,  monitoring becomes an essential activity to ensure continuity and quality of service~\cite{bi_tech}. To this end, systems and applications produce log records, that provides valuable information about the system or application health. For example, system logs serve to detect high memory usage or swapping activity in a given node, which is useful to indicate a possible anomaly in the node's behavior. 

In this paper we focus on log collection and processing for large IT infrastructures. The distinguishing features of such use case, compared to generic data collection systems (Hadoop\footnote{\url{http://hadoop.apache.org/}}~\cite{hadoop}, HBase\footnote{\url{https://hbase.apache.org/}}~\cite{hbase}, Elasticsearch\footnote{\url{https://www.elastic.co/products/elasticsearch}}~\cite{master_elastic}, etc.~\cite{bigtable}), are twofold. First, logs are produced by a large number of servers and routers and the resulting amount of data is huge. They are also produced at a very high rate, which poses a severe challenge for log collection. Second, the task of log inspection encompasses two phases that come in sequence: high-level browsing and, then, detailed analysis of selected parts of the log. Actually, in the first place, a high-level data summary must be provided; for instance, a graph showing different HTTP response codes versus time. Whenever an anomaly occurs, such as a sudden increase in the number of internal server error codes (500), the manager should be able to drill down into the data to find the root cause. However, \textit{this second phase only applies to the log data within the time interval when the anomaly occurred}. Such two-phase nature of log data examination calls for a distributed data processing and storage system specifically tailored to this use case, which is the focus of our research.

Actually, in this paper we propose a novel log processing and storage system that we call {\em Loginson} which is specifically targeted towards processing, storing and visualizing massive amounts of logs. As it turns out, detecting incidents within millions of log messages is a very challenging task, that requires a high performance system providing the above-mentioned two-phase browsing capabilities, namely high-level and detailed or drilled-down. At this point, before we provide further insights into the {\em Loginson architecture}, let us review the motivation behind this research: providing the network and system manager with a powerful, scalable and easy-to-use system which is good to tackle the daunting task of log analysis.

\subsection{Motivation}
\label{subsec:mot}
The current Internet will likely grow~\cite{ref:oecd}\cite{cisco_forecast}\cite{MS_growth} and evolve into the so-called Internet~of~Things~\cite{iot_thing}\cite{iot_future}\cite{iot_survey}, such that many small devices (meters, wearable devices, etc) will be producing data every second, not to mention large servers and routers. By data we refer to logs in the broad sense of the term, namely files with sequentially timestamped lines that have some interesting data for the operators. For example, the following box shows a log entry from the well-known Apache server:

\begin{center}
\shadowbox{
\begin{minipage}{0.9\columnwidth}
64.123.18.10 - - [07/Mar/2015:16:10:02 -0800] "GET /format.css HTTP/1.1" 200 6291
\end{minipage}
}
\end{center}

In the latter example, we can easily recognize the timestamp (date and time), IP address of the server, response code (200 OK) and response time (6291 microseconds). However, we note the log line \textit{must be parsed in order to extract such values}.

As for the log capture and processing workload, both the volume and the rate at which logs are produced are important. We set our objective to the ambitious goal of handling \textit{several millions of lines of log per second}. For instance, if log lines are 300 bytes long then a 10 Gbps link will be fully saturated at a rate of roughly 3 million logs per second. That is the offered rate of a population of 30 million small devices that generate a log line every ten seconds or a thousand servers that generate three  thousand lines of logs per second each. Such a vast workload, far beyond the state of the art as we will show, must be stored, filtered, aggregated and visualized for the purpose of detecting unexpected events or incidents. 

In what follows we describe the main motivation behind this paper, i.e. the requirements that network and system managers have for log analysis. We have gained this experience through many different projects with large banking datacenters and operators, which have been carried out in our spin-off company \textit{Naudit}\footnote{\url{http://www.naudit.es}}. Precisely, this work is motivated by many use cases in large IT infrastructures with a massive amount of logs being produced, which could hardly be managed with traditional techniques. 

\textit{The first requirement is that network managers must centralise the logs} generated by systems currently in production. We note that this is not only an operational requirement but a legal one in the banking sector\footnote{\url{https://www.pcisecuritystandards.org/}}. In order to send the logs from the servers or routers, Syslog \cite{syslog_rfc} constitutes a de facto standard, either through UDP, TCP or TLS. Therefore, we believe that Syslog is the best option due to its widespread implantation, which will continue to grow in the future. Consequently, in the central log collection repository, a {\em Loginson} component must capture the log stream at the targeted rate of millions of log lines per second. Such component provides load-balancing by splitting the data stream  as fast as possible to the different  processing and  database nodes, in round-robin fashion.  

\textit{The second requirement is to pre-process the raw data and provide a high-level summarized data overview} whereby pre-processing implies filtering, aggregating and, in general, performing a pre-analysis of the raw log data. This third hurdle has the extra handicap of dealing with the specific characteristics of each kind of log message and its format. Hence, each log type demands a different filter or aggregator (for example number of HTTP Internal Server Error codes -500- every five minutes). Furthermore, flexibility is a key requirement in the {\em Loginson} pre-processing component. The output of this component is a summarized version of the raw data that is manageable for plotting in charts, time series and histograms and offers a high-level overview which is useful to detect incidents.

\textit{The third requirement is to provide a graphical representation of the summarized data} in a user-friendly interface that allows to visualize incidents. We note that the data trimming and aggregation of the previous paragraph largely reduces the dataset to be plotted and paves the way for using open-source dashboard visualization solutions (such as \textit{Kibana})~\footnote{\url{https://www.elastic.co/products/kibana}} with modest resources (small number of virtual machines). Therefore we decided to incorporate \textit{Kibana} (with the \textit{Elastic Stack}\footnote{\url{https://www.elastic.co/v5}}) as a visualization tool rather than providing our own dashboard implementation tool. 

\subsection{Novelty}
\label{subsec:novel}
In this paper, we are specifically considering the case of real-time log monitoring in large IT infrastructures. We note that monitoring is cornerstone for networks and datacenters and, consequently, the proposed use case covers a widespread demand for monitoring at higher rates. As it turns out, monitoring demands log ingestion, processing and visualization systems whose underlying database sacrifices classical database functionalities for the sake of speed, centralization and adaptation to dashboard implementation tools. 

First of all, syslog has became a de-facto standard for sending the logs at the server side. The introduction of software agents at the server, which could perform log pre-processing and distribution to many log collection nodes, is costly and in many cases not feasible for critical servers. Such requirement impedes the use of distributed database systems in which the data producers send data to the consumer nodes by means of smart load balancing algorithms~\cite{druid}, instead of using a central point for data collection. Actually, making all servers send the logs to a central point is much simpler, because all servers use the same destination IP address, namely that of the central collection point. Such scheme does not preclude high availability, as the central point can be implemented in several hosts. Remarkably, the central collection system is stateless, which simplifies matters for redundancy.  

Second, the queries performed by system and network managers are much simpler than standard SQL statements. Whenever an incident happens, the queries that are typically performed consist of simple IP/port range searches within a given time interval (that of the incident), not complex SQL statements. While classical databases provide a better query language capabilities, our use case sacrifices query language capabilities for speed, analysis capabilities~\cite{BG_pathologies} and better data visualization features. We also note that the recent log messages are the most important and will be most queried, whilst the old ones are less and less relevant as time goes by. 

Third, we stress the importance of data visualization for system and network monitoring. With many different servers and network segments to monitor, system and network managers strive for building a simplified, yet complete, graphical dashboard of the datacenter. Such a requirement calls for online data summarization capabilities, as all the data cannot be displayed in a dashboard.

{\em Loginson}'s novelty lies in tackling all the above requirements at a very high log ingestion speeds. Thus, {\em Loginson} differs from more complex distributed database systems that are targeted to more generic data collection and search use cases, providing a novel ad-hoc system for log collection, processing, storage and visualization.

\begin{table}[!b]
\centering
\caption{System Comparison Summary}
\label{tab:summary}
\begin{threeparttable}
\begin{tabular}{|c|c|c|}
\hline
Purpose                                                                                                                            & System                                                  & Performance                                                                                  \\ \hline
\multirow{5}{*}{\begin{tabular}[c]{@{}c@{}}\\\\Log\\ Message\\ Brokers\end{tabular}}                                                   & Syslog-ng                                               & 650K log/s multithread ~\cite{syslogng}                                                                       \\ \cline{2-3} 
                                                                                                                                   & \begin{tabular}[c]{@{}c@{}}Apache \\ Kafka\end{tabular} & \begin{tabular}[c]{@{}c@{}}300K log/s \\ Using one producer \\ and one consumer\end{tabular} \\ \cline{2-3} 
                                                                                                                                   & Apache Flume                                            & 0.77M log/s/thread                                                                           \\ \cline{2-3} 
                                                                                                                                   & Logstash                                                & 29.5K log/s/thread                                                                           \\ \cline{2-3} 
                                                                                                                                   & FluentD                                                 & 13K log/s/thread                                                                             \\ \hline
\multirow{2}{*}{\begin{tabular}[c]{@{}c@{}}Database\\ Storage\end{tabular}}                                                        & Cassandra                                                & 251K TPS\tnote{1} on a single node ~\cite{scylla:benchmark:one}                                                                   \\ \cline{2-3} 
                                                                                                                                   & ScyllaDB                                                & 1.8M TPS on a single node ~\cite{scylla:benchmark:one}                                                  \\ \hline
\multirow{4}{*}{\begin{tabular}[c]{@{}c@{}}Pre-process \\ (Split and join \\ the log fields)\end{tabular}} & \begin{tabular}[c]{@{}c@{}}Apache \\ Storm\end{tabular} & \begin{tabular}[c]{@{}c@{}}900K logs/s \\ Using 100\% of all cores\end{tabular}              \\ \cline{2-3} 
                                                                                                                                   & AWK                                                     & 900K log/s/thread                                                                              \\ \cline{2-3} 
                                                                                                                                   & Python                                                  & 500K log/s/thread                                                                              \\ \cline{2-3} 
                                                                                                                                   & Perl / Ruby                                             & 300K log/s/thread                                                                              \\ \hline
\multirow{7}{*}{\begin{tabular}[c]{@{}c@{}}Graphical \\ Representation\end{tabular}} & Kibana                                                  & \begin{tabular}[c]{@{}c@{}}Official Elastic tool, like ES\tnote{2} \\ Proper dashboard composer tool \\ User friendly and chart versatility \end{tabular}                                                                              \\ \cline{2-3} 
                             & Grafana                                                  & \begin{tabular}[c]{@{}c@{}}Unofficial, lacks support \\ Proper dashboard composer tool \\ Charts almost restricted to time series\end{tabular}                                                                              \\ \cline{2-3}                                                                                                      & D3.js                                             & \begin{tabular}[c]{@{}c@{}}Requires complex middleware \\ Big catalog of charts \\ JS\tnote{3}~~library, requires custom development\end{tabular}                                                                              \\ \hline
\end{tabular}
\begin{tablenotes}
\item[1] Transactions per second
\item[2] Elasticsearch
\item[3] JavaScript
\end{tablenotes}
\end{threeparttable}
\end{table}

\section{State of the Art}
\label{sec:soa}

Most of the distributed processing and storage systems are useful for generic data analysis scenarios, not specifically for log collection, processing and visualization. Their main goal is to perform complex analysis of a large quantity of persistent data~\cite{BG_problem} that is inserted at a relatively low rate. 

Actually, traditional big data systems can scale horizontally~\cite{netflix:cassandra} to a large number of nodes, which are load-balanced through complex algorithms. Such systems try to hide low-level complexity from the programmers, allowing them to focus on the problem to solve. The tradeoff is that this abstraction adds a noticeable overhead when using fewer nodes, as vertical scalability is sacrificed to achieve simplicity.

In the case of {\em Loginson}, the loss of vertical scalability is not acceptable. We have soft real-time constraints that cannot be met by any of the generic systems we have tested unless a very large number of nodes is used\cite{mit:achieving}. Furthermore, vertical scalability makes it feasible to use active-active configuration for data replication, because the number of nodes involved is not so high. Our conclusion is that a custom system is necessary for the real-time analysis and storage components of {\em Loginson}. After the amount of data is reduced by the former, traditional systems like \textit{Elasticsearch} can be used to offer high level analytics.

In what follows we will review the state-of-art for our specific use case of log storage and visualization to conclude that no system to date fulfills the requirements set forth in the previous section. We divide the state of the art in four sections: message brokering and load balancing systems, database nodes, pre-processing (data transformation) components and graphical representation. As a summary, Table \ref{tab:summary} presents the most salient performance aspects of the different tools included in the state of art.

\subsection{Message brokering and load balancing for log centralization}
\label{subsec:brokering}

The most straightforward choice for log centralization is to adopt one of the syslog daemons already existing in the open-source community. Out of these, \textbf{Syslog-ng} is the fastest, because it was designed for  speed. It can be easily configured to receive logs from a client syslog daemon and to store them in a centralized location, which make it the ideal solution for our problem. Unfortunately, \textbf{Syslog-ng} cannot achieve a throughput of three million log lines per second, peaking at 650,000~\cite{syslogng}.

Another popular system that comes at hand to centralize logs is called \textbf{Apache Kafka}~\cite{kafka_intro}, a broker that can receive data from multiple sources and distribute it to several destinations. Kafka uses a publish-subscribe pattern which allows to persist the data in disk as well as to distribute it to the subscribed consumers (logstash, fluentd, TCP clients, etc.). Kafka categorizes messages by topics, making it possible for consumers to subscribe to one or many topics. Such topics are divided into a number of partitions, which can be accessed in parallel by multiple consumers. In the case of producers, they can publish their log messages to a specific topic.

In benchmarks published by Linkedin~\cite{linkedin:kafka}, a three node cluster is capable of ingesting 2 million events/s generated by three producers. The tests also show that producer throughput is not reduced by adding consumers, as one producer is capable of inserting 800,000 events/s with or without consumers.

Unlike Syslog-ng, Kafka is not focused on log processing and additional work becomes necessary. Kafka seemed to meet the needs of our use case, so we tested it using a server with two \texttt{Intel Xeon E5-2630 v2 @ 2.60GHz} CPUs, 32GB of RAM and a RAID 0 with 10 HDD. We found that Kafka did not satisfy our requirements, it used 100\% of the machine CPU, showing an ingestion rate of 300,000 log entries (size 117 bytes) per second using one producer while reading from one consumer. Other log collection systems\footnote{\url{https://www.loggly.com/}} that use Kafka also get similar results. This throughput is an order of magnitude below our requirements, so we had to discard the use of Kafka in {\em Loginson}.

\begin{figure}[!th]
\begin{center}
\includegraphics[width=0.9\columnwidth]{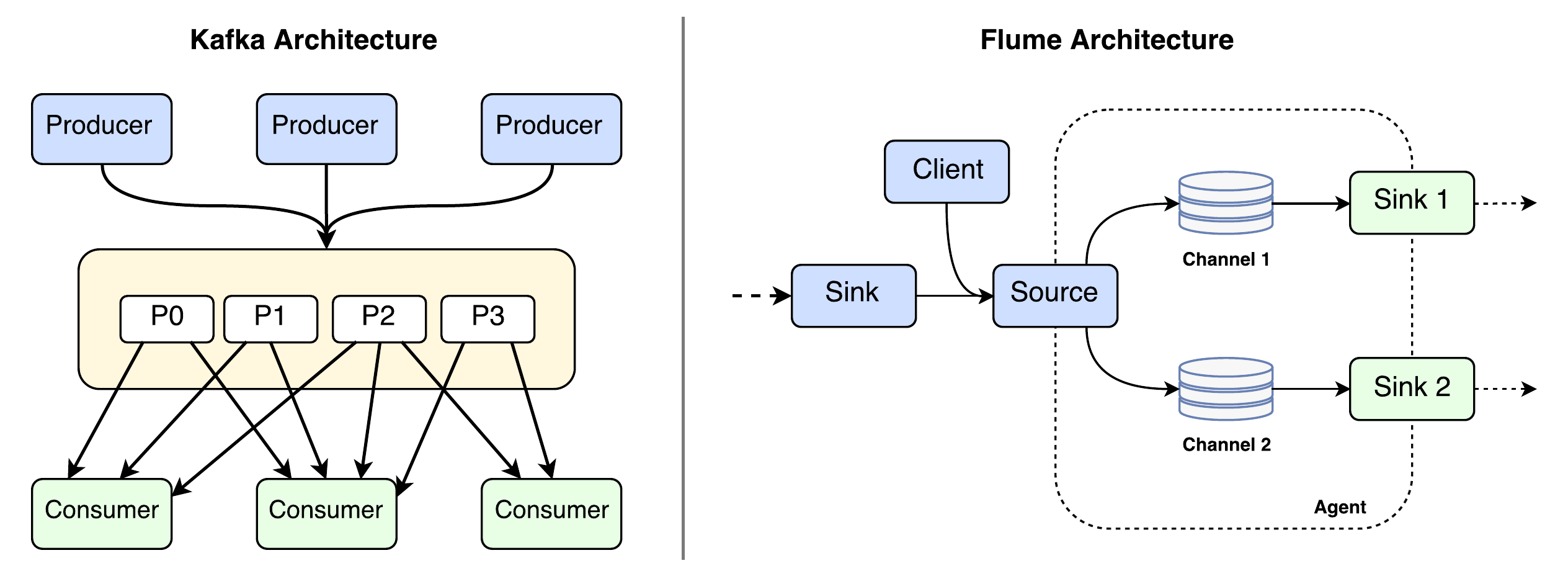}
\caption{Flume and Kafka architectures}
\label{dia:flume_kafka_arch}
\end{center}
\end{figure}

An alternative to Apache Kafka is \textbf{Apache Flume}\footnote{\url{https://flume.apache.org/}}, with the main difference being that Flume pushes the data to the destination, whilst Kafka's clients pull the data from Kafka's queues at their own pace. While Flume's sinks can read only from one channel, multiple sinks can be read from the same channel. Furthermore, multiple sources can write on the same channel. Figure~\ref{dia:flume_kafka_arch} compares both architectures. As Kafka's subscribers, Flume's sinks can be of different types, like HDFS\cite{hdfs}, Elasticsearch, HBase, etc.

We tested Flume on the same machine with two different sink configurations: Null Sink, that discards all the events it receives from the channel\cite{flume_book}; and File Roll Sink that stores the events in files. To achieve maximum performance the output files were stored in a RAM disk. Flume achieved a performance of 1,500,000 log messages per second and thread with the former kind of sink and 770,000 logs per second and thread, with the latter using a log size of 291 bytes. More details about the test can be found on Figure~\ref{fig:loginson_flume} in Section~\ref{sec:results}. Again, this alternative did not match our performance needs.

There are other systems dedicated to log processing that are worth mentioning even if they do not provide the necessary throughput. \textbf{Logstash}\footnote{\url{https://www.elastic.co/products/logstash}}, developed by \textit{Elastic}, collects, processes and forwards log messages from and to different systems (TCP, UDP, \textit{Elasticsearch}, \textit{Kafka}, \textit{0mq}, etc.) promising 50,000 events per second\footnote{\url{https://www.elastic.co/blog/logstash-1-5-0-ga-released}} when using grok filters for log patterns.

Similarly, Treasure Data developed \textbf{FluentD}\footnote{\url{http://www.fluentd.org/}}, an alternative to Logstash that presumes of a throughput of 800,000 events per second~\cite{fluentd:blog} in their Big Data cloud service, but without specifying any details~\cite{fluentd:slides}. On their website they describe the system as being able to process 13,000 events/second/core~\cite{fluentd:architecture}.
\begin{figure}[!tbh]
\begin{center}
\includegraphics[width=\columnwidth]{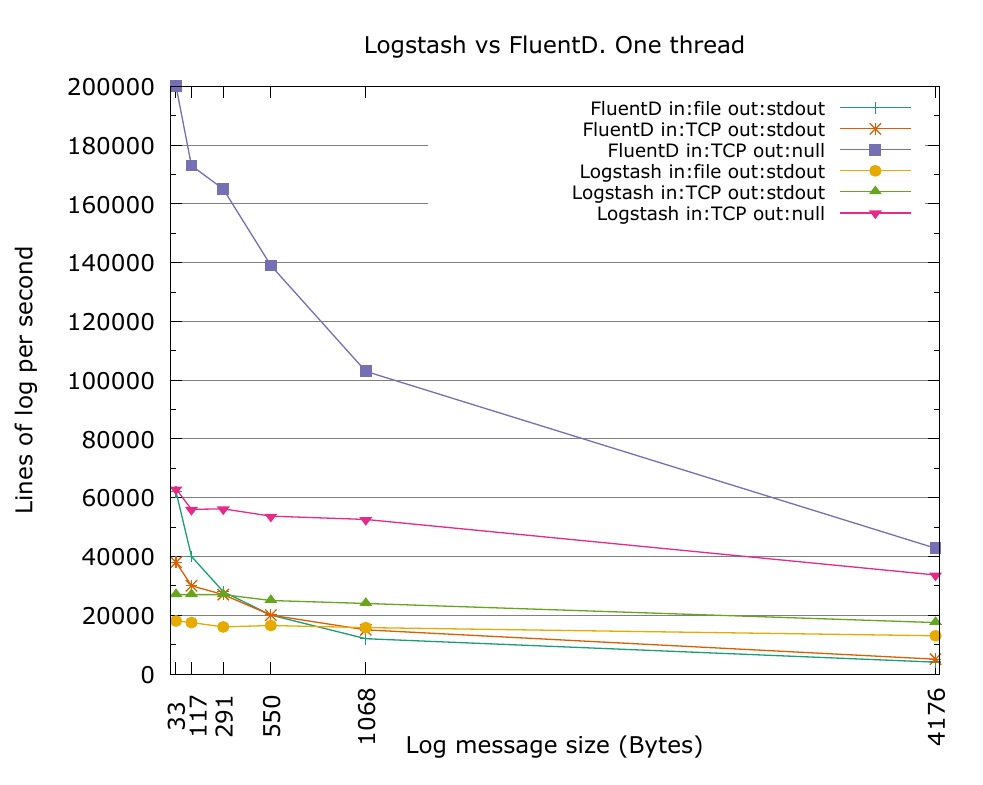}
\caption{Logstash and FluentD benchmarks}
\label{fig:logstash_fluentd_benchmark}
\end{center}
\end{figure}
We tested both Logstash and FluentD on the same machine as the previous tests, without performing any process or transformation on the log entries but the default, which is to add a timestamp to the log, with a first test forwarding messages to stdout after being read from files on a RAM disk; a second test in which the log entries are received via TCP; and a last test also getting the log lines via TCP but then sent to null using their corresponding null output plugins. As seen in the Figure~\ref{fig:logstash_fluentd_benchmark}, both Logstash and FluentD do not surpass 60,000 log messages per second except for the tests in which log messages are sent to null. Hence, we had to dismiss such systems.

\subsection{Database nodes}
\label{subsec:dbnodes}
After receiving the log messages in a load-balancer they must be split among several database/storage nodes. That way the write workload per disk is reduced, which leaves room for the read operations not to affect the write performance.

There are several options to store this data, from big data systems to simple key-value storage. Our extensive tests have shown that, while \textbf{our solution needs only one server to process up to 10 million lines of log per second}, big data storage solutions like \textbf{Apache's HDFS} or \textbf{Cassandra}\footnote{\url{http://cassandra.apache.org/}} are only capable of meeting our targeted input rate by using at least 5 servers, so we had to discard them for cost-effectiveness~\cite{nosqlbench}. Such systems are better tailored to a more generic use case at the expense of performance, with features like fault-tolerance and high availability that, despite its value, can be achieved by other means such as system redundancy. 

In the case of key-value storage, there are several benchmarks~\cite{keyvaluebench} that show that none of them can achieve a write speed fast enough for our purposes. Moreover, such systems are designed for random reads and writes, so they were discarded in favor of sequential flat file storage, as log entries are usually accessed in sequential chunks.

Some systems like \textbf{ScyllaDB}\footnote{\url{http://www.scylladb.com/}}~\cite{scylla:architecture} (an open-source C++ rewrite of Cassandra) claim to process up to 1M transactions per second per server. One of their benchmarks~\cite{scylla:benchmark:cluster} uses a cluster with 3 nodes with 128GB of RAM and 4x 960GB SSDs on RAID each. Using a replication factor of 3, the cluster provides an average performance of 1.9M transactions per second (TPS).

Another benchmark has been reported in~\cite{scylla:benchmark:one} using a single node which gives a throughput of 1.8M TPS. While the performance of ScyllaDB surpasses Cassandra's (251K TPS for the latter test), the tests of these benchmarks were done using SSDs disks, which have a better performance on simultaneous write/read operations than the conventional HDDs at a higher cost. For example, a system that collects 3 millions log lines per second produces an approximate 42 TB worth of log messages per day, so using SSDs would be extremely expensive.

Other solutions such as \textbf{MapRDB}\footnote{https://www.mapr.com/} have a commercial license of 4,000\$ per node (2011)~\cite{mapr:cost}. MapR accomplish the ingest of 100,000,000 points per second with 4 nodes and a replication factor of 3 thanks to its own filesystem mapR-FS~\cite{mapr:benchmark}.

\textbf{Splunk}, is a search server similar to Elasticsearch, which is available with a commercial license at a price of 28,750\$ for 10GB/day, 172,500\$ for 100GB/day on their Enterprise Version and 15,870\$ for 10GB/day for their cloud service. It is a closed-source software which, according to their own tests~\cite{splunk_test}~\cite{splunk_performance}, gives a performance of 80,000 events per second on a Linux 64-bit HP DL380G7 machine with 2x6 Xeon @2.67 Ghz and 12GB of RAM. \textbf{Elasticsearch}, which, contrary to Splunk, is an open-source solution, achieves the same speed with a dual Xeon X2699 @2.3 Ghz and 256GB of RAM using just a 4GB heap, according to their benchmarks \cite{elastic_performance}.

\subsection{Pre-processing the logs}
\label{subsec:preprologs}
To pre-process the raw data, a real time streaming system is necessary. A popular solution is \textbf{Apache Storm}\footnote{\url{http://storm.apache.org/}}, a distributed realtime computation system. We ran a benchmark in one node with two 6 core Intel Xeon E5-2630 v2 CPUs. Using a 100\% of all cores, storm is capable of transferring 900,000 log lines per second, just splitting and joining the fields of the log line. Other systems like \textbf{Apache Spark}\footnote{\url{http://spark.apache.org/}} streaming and \textbf{Apache Flink}\footnote{\url{https://flink.apache.org/}} had worse performance.

Such performance figure of 900,000 log lines per second was obtained using the AWK programming language, and with a single core. Furthermore, such throughput decreases to 500,000 log entries per second when using Python and to 300,000 with Ruby or Perl.

Languages like AWK or Python can be linked using small wrappers or unix pipes to create a streaming pipeline. They offer the benefit of being easy to use and well-known, with no need of specialized programmers.

\subsubsection{Serialization}
\label{subsec:serialize}
The problem of serialization arises when  parsing at high speed becomes necessary to convert log messages from its original format to \textit{JavaScript Object Notation}, JSON. One of the main constraints we work with is the hindrance of not to incorporate any software to the log senders as they are usually systems in production that should not be touched. As a consequence, the log format cannot be altered in the source even though some tools like Apache, Nginx or even syslog-ng OSE (which is able to send logs directly to \textit{Elasticsearch})~\cite{syslogng:elastic} support \textit{JSON logging}.

As explained previously, systems like {\em Logstash} or {\em FluentD} are not fast enough for collecting, nor parsing, the log messages. The resulting conversion throughput is between 15,000 and 30,000 log lines per second depending on the log size.
\subsection{Graphical representation}
\label{subsec:graphical}
The visualization system is an essential part of the architecture as it interacts with the final user of the system, hence, the simplicity and flexibility are of fundamental importance to favor user experience. In this regard, charts must be dynamic and flexible enough to meet the needs of the data analysts, yet quick enough to get an overview of the data in a matter of seconds. 

Some tools like \textbf{D3.js}\footnote{\url{https://d3js.org}} (a JavaScript library) feature a huge catalogue of different kind of charts (tree maps, box plots, bubble charts, etc.) but require a web developer with good programming skills in order to make any change to the charts, in addition to the fact that they are not able to connect by themselves to the datastore. Hence, a complex middleware must be put together to make them usable. 

Moreover, tools like \textbf{Grafana}\footnote{\url{http://grafana.org}} or \textbf{Kibana} are able to connect to the source of the data and display a large variety of charts without requiring advanced programming skills. In the case of Grafana, it lacks variety of charts, and it is restricted to time series only. 

Finally, \textit{Kibana} has both a variety of charts (pie charts, time series, histograms, etc.) and a data browsing API to the \textit{Elasticsearch}\footnote{\url{https://www.elastic.co/products/elasticsearch}} indexes which makes it straightforward to manage or change the chart dashboards. Furthermore, \textit{Kibana} and \textit{Elasticsearch} are in constant development by the \textit{Elastic team}, which recently released \textit{Timelion}\footnote{\url{https://www.elastic.co/blog/timelion-timeline}}, an specific plugin for \textit{Kibana} that provides time series functionality for the charts (derivatives, moving averages, etc.), thus adding support for some functionalities just seen on \textit{Grafana}. \textit{Kibana}, unlike \textit{Grafana}, also supports embedding dashboards and charts allowing for custom web pages. 

All the above-mentioned features makes \textit{Kibana} our graphical tool of choice for plotting the charts and graphs. 

\section{Our architecture}
\label{sec:arch}
In {\em Loginson} we receive all logs in a centralized server called \textit{LogFeeder} that adds information like the timestamp to the logs, to avoid timing problems with logs from different timezones or from desynchronized sources. Then, they are split between several storage nodes,  using round robin. The storage consists of flat files of 1 GB size. Such size is large enough for  the writes and reads to take advantage of high speed sequential operations on spinning disks. If high availability is required, an active-active {\em Loginson} system with data replication could be used to avoid loss of data. Such data replication comes at a modest cost, because nodes are highly utilized (vertical scalability), in contrast to the low utilization of pure horizontally scalable systems. 

We note that the proposed system provides high performance due to its simplicity and vertical scalability. If higher performance is required, more storage nodes can be added to split the load. The central collection point {\em LogFeeder} becomes a bottleneck, though, which can be alleviated by the use of load balancer and several {\em LogFeeder} nodes. In any case, {\em LogFeeder's} throughput reaches several million log records per second (see figure \ref{fig:logfeeder_speed1}), which is sufficient for extremely large IT infrastructures.

\begin{figure*}[!tb]
\begin{center}
\includegraphics[width=0.9\columnwidth]{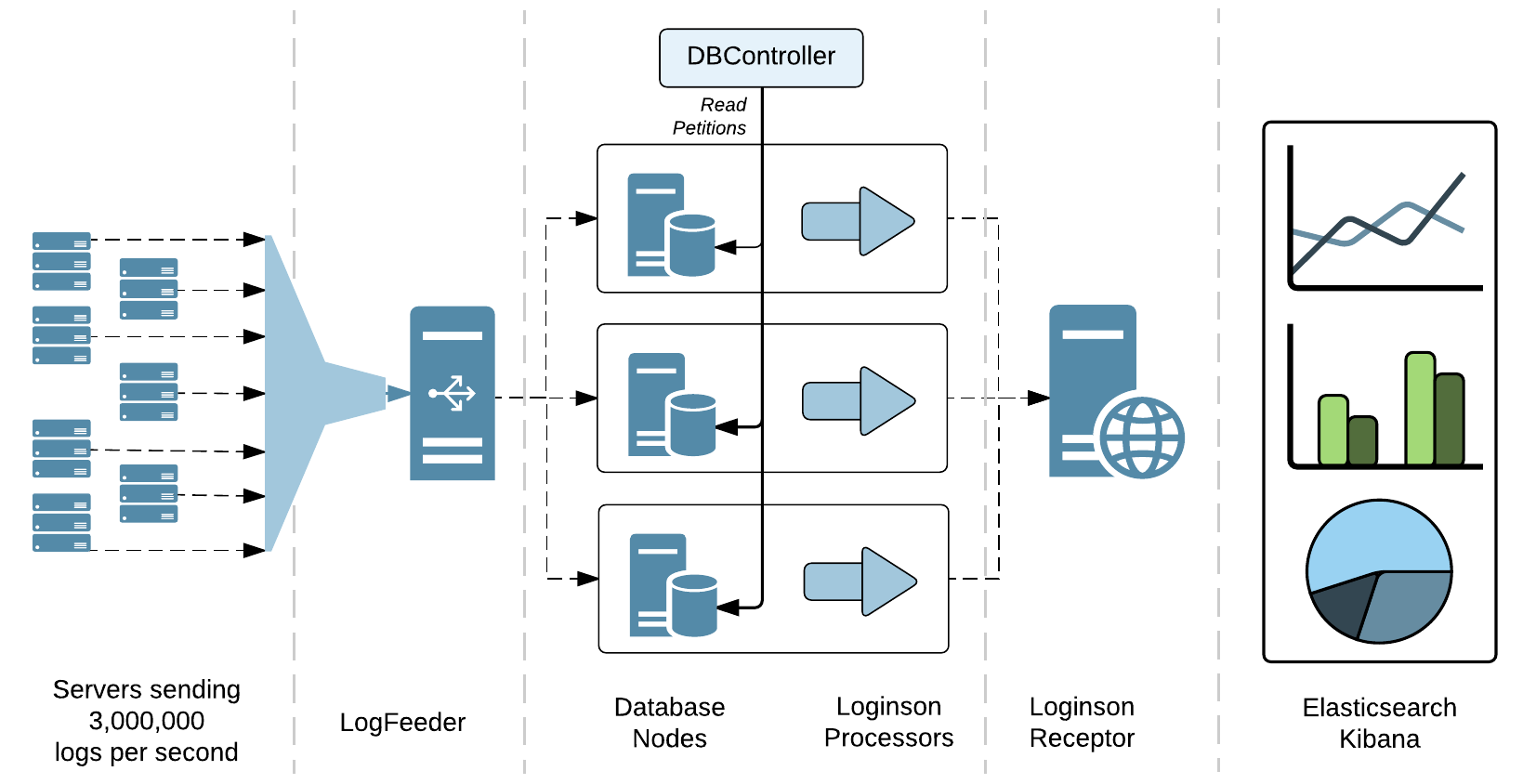}
\caption{Loginson architecture}
\label{fig:arch}
\end{center}
\end{figure*}
While the logs are stored on the database nodes, a summary is constructed using the programming language AWK (for instance, moving averages of the data), and sent to a component called {\em LoginsonReceptor}. Such component centralizes the processed logs and inserts them in an \textit{Elasticsearch database} so \textit{Kibana} can be used to visualize different metrics on the summarized data. Usually, {\em LoginsonReceptor} could be placed on a different node, such that the access to the summarized data does not interfere with the collection of new data. To further drill down into the data, our solution exposes a web API to access the logs stored in the database nodes. Such component, called \textit{DBController}, is capable of queueing several read requests that can be deleted at any moment. Thus, the user is able to abort and modify the drill-down requests without waiting for them to finish. The logs are then processed in the same way as in real time and sent to the {\em LoginsonReceptor}.

In what follows, we will provide detailed insight of each of the system components. Figure \ref{fig:arch} shows the {\em Loginson} architecture.

\subsection{Message brokering and load balancing for log centralization}
\label{subsec:arch_brokering}
We choose to perform load balancing from a central node to the database nodes instead of making the clients send the logs directly to specific database nodes. By doing so, we avoid that some of the database nodes receive more load than others and simplify configuration at the client side.
Such load balancing node is called \textit{LogFeeder}. It receives log lines (within UDP) using the high speed HPCAP network driver~\cite{Moreno12} for Intel\textsuperscript{\textregistered} Ethernet 10 Gb PCI Express NICs, capable of receiving network traffic at 10~Gbps. 

\begin{figure}[!b]
\begin{center}
\includegraphics[width=0.8\columnwidth]{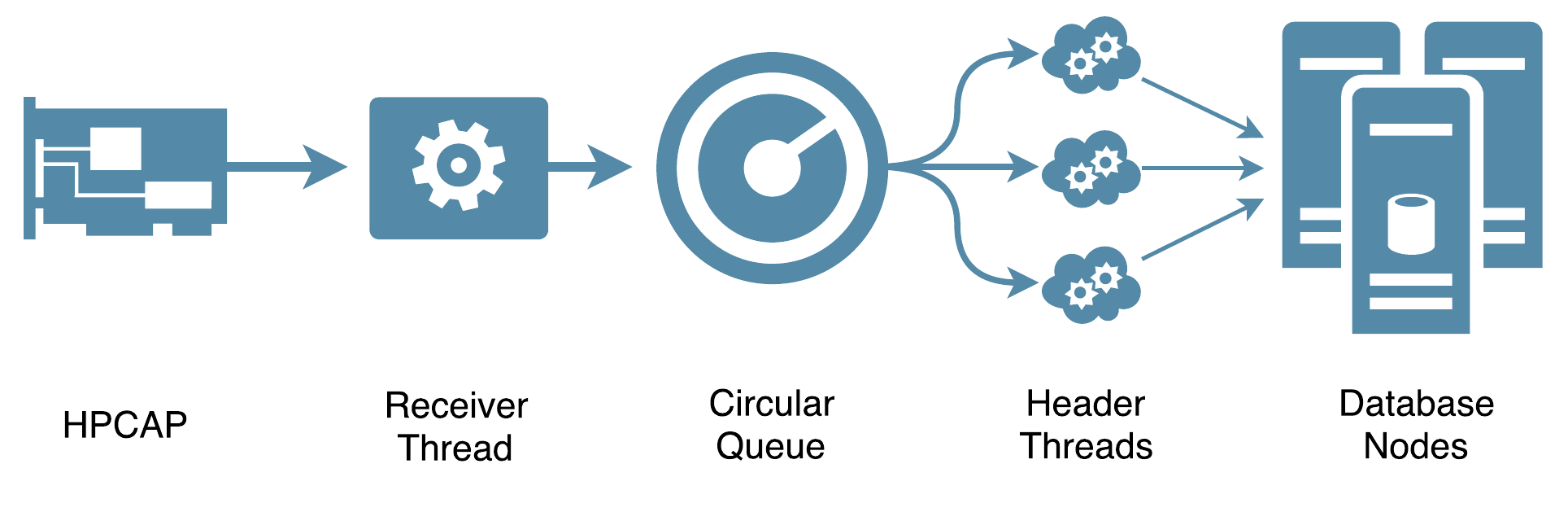}
\caption{LogFeeder}
\label{fig:logfeeder}
\end{center}
\end{figure}

To achieve high speed, \textit{LogFeeder}, as shown in figure~\ref{fig:logfeeder}, uses several threads that share information through a circular queue of buffers. This queue is protected through a mutex. Buffers are used to avoid locking the mutex for each log, which would decrease performance dramatically.

One of the threads, called \textit{Receiver Thread}, reads the logs from the network interface, performs timestamping and writes them to a buffer in the circular queue. This buffer is then made available for processing. and the Receiver Thread goes on to the next buffer for writing. The rest of the threads, called \textit{Header Threads}, read a filled buffer from the circular queue in a first-come first-serve basis. For each line of log in the buffer, such threads add a header with information about the log source and send them to the database nodes using a Round-Robin approach.

An advantage of using Round-Robin instead of more complex algorithms is that logs are uniformly split between the database nodes. This makes subsequent access to the database faster, as it can take advantage of data parallelism.
\subsection{Database nodes}
\label{subsec:arch_storage}
Logs are ordered by timestamp and will be accessed in batches for analysis. Most existing databases are not adequate for the job, as they are optimized for random reads and writes. Instead, each storage node works with sequential flat files. Within such files, logs are preceded by a 64 bytes header that stores the timestamp and type of log.

\subsubsection{Writing to the database}
\label{subsec:writedb}

Logs are received using TCP from the \textit{LogFeeder} and copied to one of several buffers in a pool. When the buffer is full, it is written to disk asynchronously as a single file. This way the database does not block and continues receiving logs and copying them to another buffer. As mentioned before, the size of these buffers is 1 GB to take advantage of fast sequential writes in spinning disks.
Such files are then indexed in a SQL database by storing the first and last timestamps. The former index will be used to get a list of files whenever logs need to be accessed. While this thread is receiving logs, another one is processing them as will be explained on Section~\ref{sec:preprocess}.
\begin{figure}[!b]
\begin{center}
\includegraphics[width=\columnwidth]{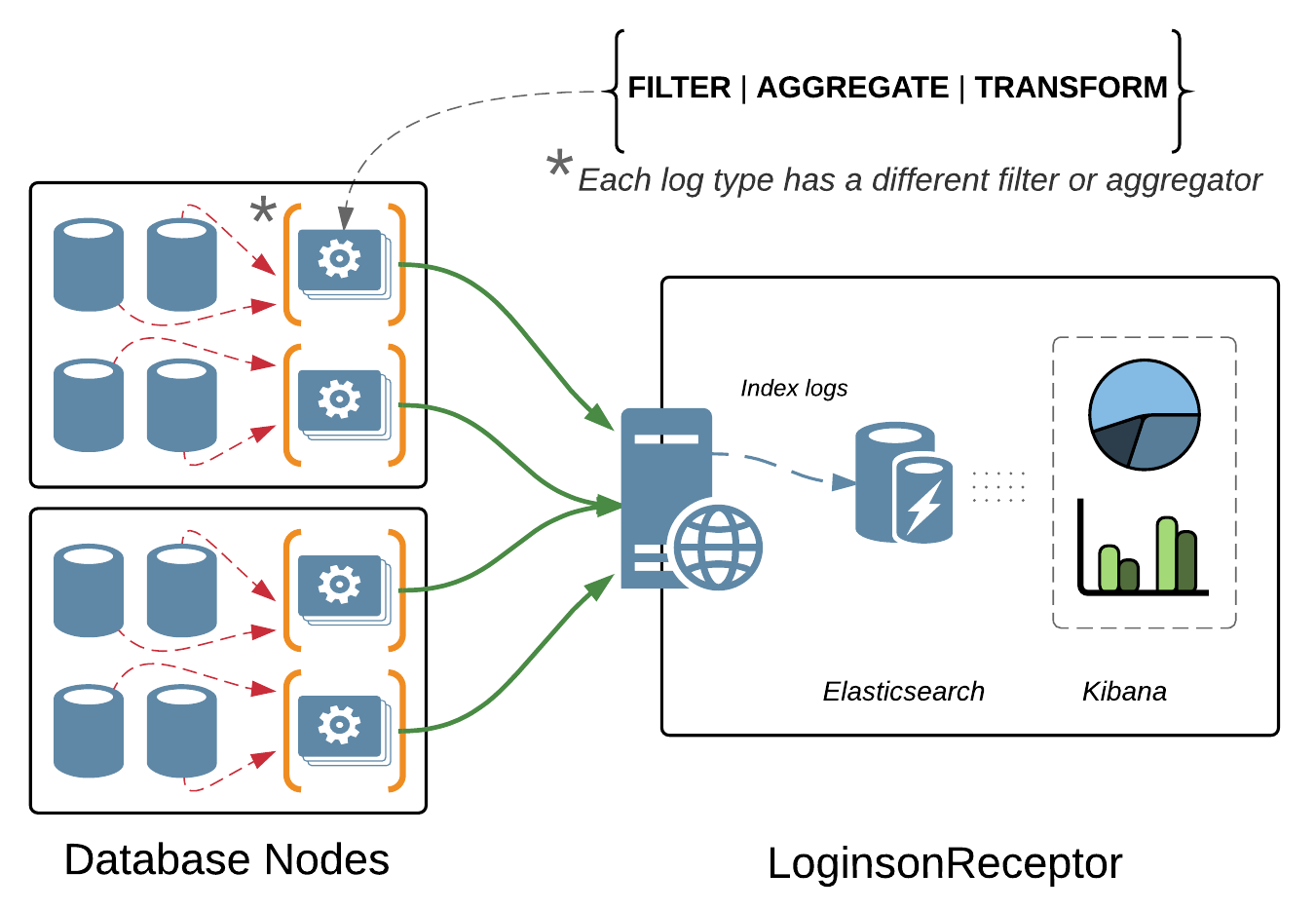}
\caption{LoginsonReceptor}
\label{fig:loginsonReceptor}
\end{center}
\end{figure}
\subsubsection{Reading logs}
\label{subsec:readlogs}
The logs are split between several nodes, so we need a single point of access that we called \textit{DBController}. It receives read queries and forwards them to each database node. Read queries are performed through an HTTP API and enqueued until the database nodes are available for reading. We note that the queries can also be canceled even when they are running.
When a database node receives a read query from the \textit{DBController}, it first searches in the SQL database for all the files containing logs between the two given timestamps. Each file is loaded to memory and only the logs of the requested type are processed and sent to a program called {\em LoginsonReceptor} for visualization.

\subsection{Pre-processing the logs}
\label{sec:preprocess}
To provide a high-level overview of the logs in real-time, a given summary analytics (for example, moving average of HTTP response time) of each kind of log message is sent to its correspondent AWK process through a UNIX pipe. Each of the database nodes use these AWK scripts, one for each type of log, which are utterly flexible allowing for filters, aggregations, or transformations of the relevant log messages as well as its JSON serialization, which is compulsory for  \textit{Elasticsearch}.

The result is then sent via TCP to {\em LoginsonReceptor} (Figure~\ref{fig:loginsonReceptor}), in the graphics visualization server, where the data of each database node is collected and indexed on \textit{Elasticsearch} for subsequent visualization on Kibana and Timelion. {\em LoginsonReceptor} manages the creation of indices and mappings on \textit{Elasticsearch} for each kind of log message making use of the \textit{Elasticsearch Java API}. AWK was chosen for its flexibility and speed as shown on Section~\ref{sec:preprocesstest}, as well as its high-level capabilities like the use of associative arrays or regular expressions.

In the next section we present a performance evaluation of each component of our architecture as well as the whole system.
\begin{figure}[!p]
\begin{center}
\includegraphics[width=0.9\columnwidth]{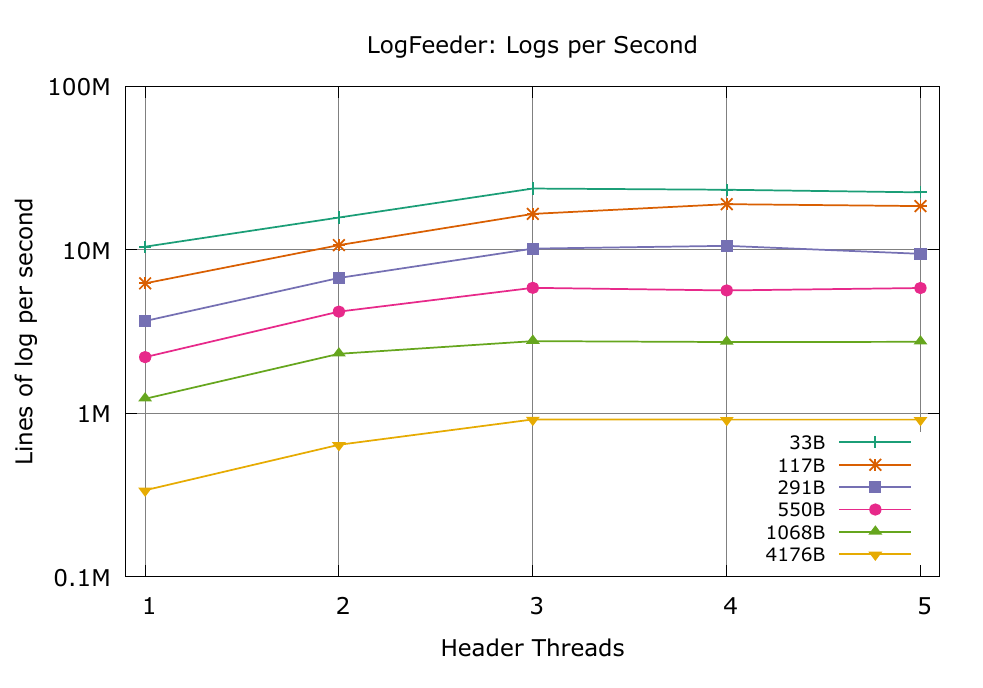}
\caption{LogFeeder Speed}
\label{fig:logfeeder_speed1}
\end{center}
\begin{center}
\includegraphics[width=0.9\columnwidth]{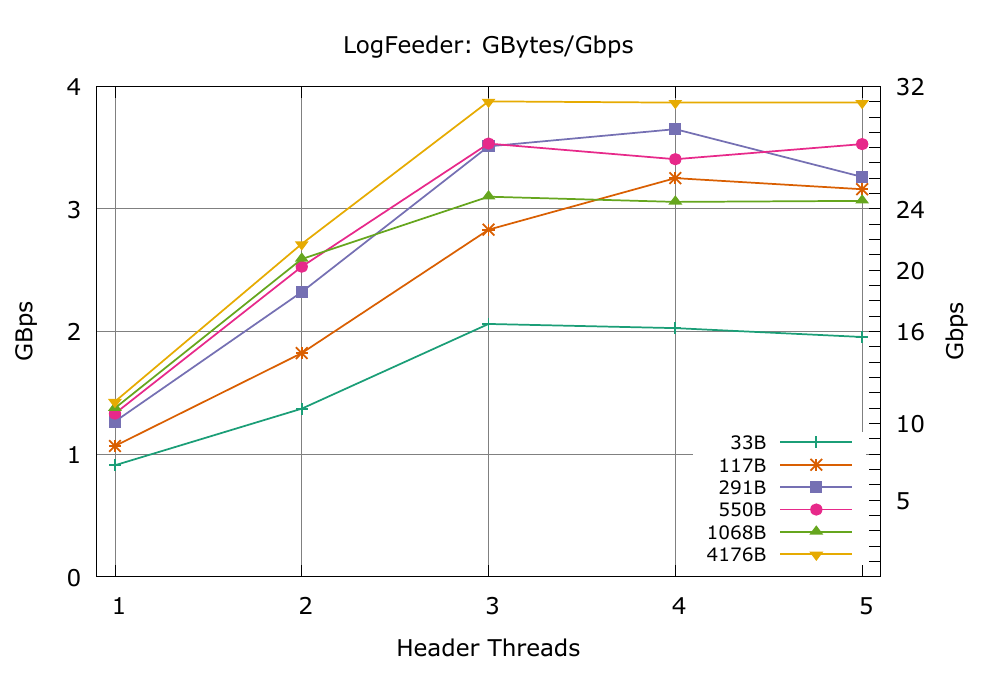}
\caption{LogFeeder Gbps/GBps}
\label{fig:logfeeder_speed2}
\end{center}
\end{figure}

\section{Performance evaluation}
\label{sec:results}
In this section we perform extensive stress testing to ensure that {\em Loginson} fulfills our performance requirements. We tested each component separately and, then, we ran tests on the whole system.

First, we studied {\em LogFeeder} performance versus  the number of threads used.  We also compared the input data with the output and checked that both were equal.
Next, we tested the database nodes and performed fine-tuning in order to achieve a constant write rate on a spinning disk. Furthermore, we noticed that log pre-processing did not affect the write performance. Again, our tests showed that the data written to disk was the same as the one inserted in {\em LogFeeder}.

Afterwards, for the pre-processing stage, we compared different options for the serialization of the log messages as it is a mandatory step before forwarding the logs to {\em LoginsonReceptor}. These three options were faster than the results obtained with Logstash and Fluentd (Figure \ref{fig:logstash_fluentd_benchmark}) but we lastly opted for AWK owing to its speed and flexibility. We also benchmarked the performance of {\em LoginsonReceptor} indexing logs on \textit{Elasticsearch} in order to check the maximum indexing speed of the latter.

Finally, two tests of the whole system were conducted. To this end,  we used four nodes, one of them running \textit{LogFeeder}, two as database nodes and another one running \textit{DBController} and \textit{Elasticsearch}. In the last test only one node with a RAID~0 was used to host \textit{LogFeeder} and the database. 

Most importantly, we note that Flume is the only system close to our requirements, thus, the section concludes with a performance comparative between Flume and {\em Loginson}, leaving aside the rest of alternatives.
\begin{table}[t]
\begin{center}
 \captionsetup{justification=centering}
 \caption {LogFeeder Performance with logs of 291 Bytes\\Number of logs passing through the system every 100 ms.}
  \label{tab:logFeeder_test}
 \begin{tabular}{| c | c | c | c |}
 \hline
 \thead{Number of Header threads} & \thead{Mean} & \thead{Median} & \thead{Standard Deviation} \\ \hline
 1    & 512600 & 516100 & 17945.03  \\ \hline
 2    & 693200 & 701900 & 25254  \\ \hline
 3    & 1038000 & 1044000 & 17588.47  \\ \hline
 4    & 1070000 & 1073000 & 30756.99  \\ \hline
 \end{tabular}
\end{center}
\end{table}

\begin{figure}[!p]
\begin{center}
\includegraphics[width=0.9\columnwidth]{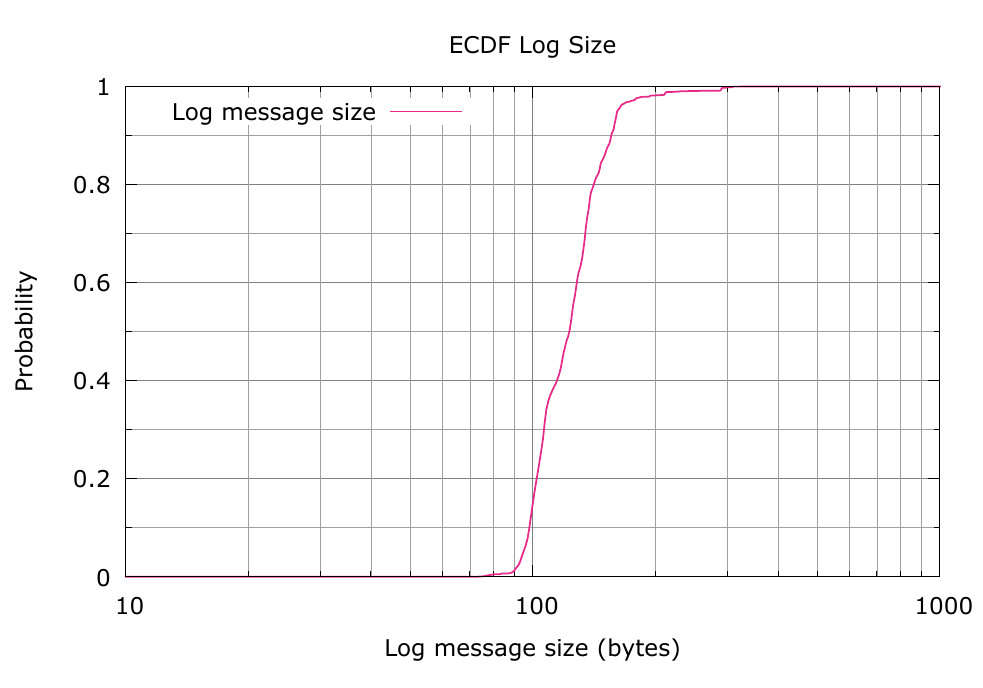}
\caption{ECDF of Log message size}
\label{fig:ecdf}
\end{center}
\begin{center}
\includegraphics[width=0.88\columnwidth]{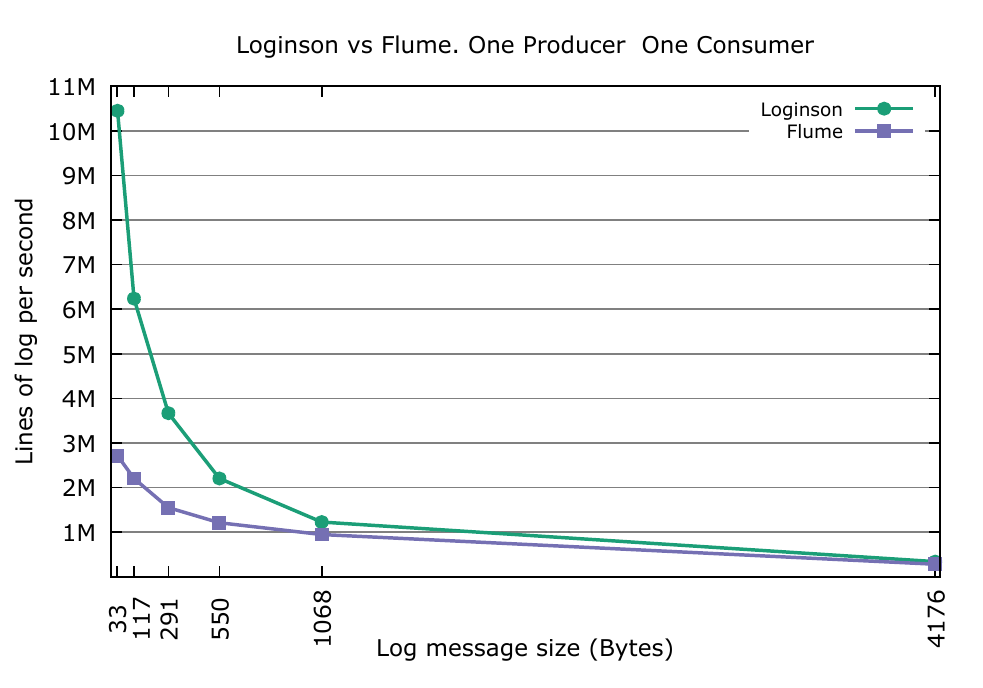}
\caption{Loginson vs Flume performance comparison}
\label{fig:loginson_flume}
\end{center}
\end{figure}

\subsection{LogFeeder: Message brokering and load balancing for log centralization}
\label{sec:logfeeder}
We tested \textit{LogFeeder} in a server with two \texttt{Intel Xeon E5-2630 v2 @ 2.60 GHz} CPUs. Each CPU featured six cores and hyperthreading was disabled. The server incorporated \texttt{32 GB} of RAM memory divided uniformly between two NUMA nodes. The server was running CentOS 7, which was configured to enforce that only the first CPU could be used, by means of \textit{taskset}. 

In our first test, we used \textit{LogFeeder} to centralize lines of log of different sizes between 33 and 4176 bytes. The obtained throughput is shown in figures \ref{fig:logfeeder_speed1} and \ref{fig:logfeeder_speed2}, as well as in Table~\ref{tab:logFeeder_test}.

We notice that for any log line size, \textit{LogFeeder} provides a throughput higher than 10~Gbps. It can also process more than 10 million log lines per second, if they are shorter than 291 bytes, which, as Figure \ref{fig:ecdf} shows, is the commonest. In order to assess what is the typical size of log lines, we studied a sample from a real production system.  We  took several days worth of Apache's logs and represented the log line sizes with the Empirical Cumulative Distribution Function (ECDF) of Figure~\ref{fig:ecdf}. Note that the median log size is around 117 bytes and that 99\% of the logs are shorter than 291 bytes, which reinforces the point that a log collection system should be able to work with small log line sizes at very high speed.

We also observe that the highest throughput is achieved when using three or four threads, depending of the log line size. Figure \ref{fig:loginson_flume} compares Apache Flume and {\em Loginson} performance. The figure shows the performance of both systems using 2 threads, one for the producer (which we previously called \textit{Receiver Thread} for {\em Loginson}) and another one for the consumer (\textit{Header Thread} for {\em Loginson}). As seen, {\em Loginson} outperforms Flume with small log message sizes, which are the most common as shown in the previous ECDF.

\begin{figure}[!h]
\begin{center}
\includegraphics[width=0.9\columnwidth]{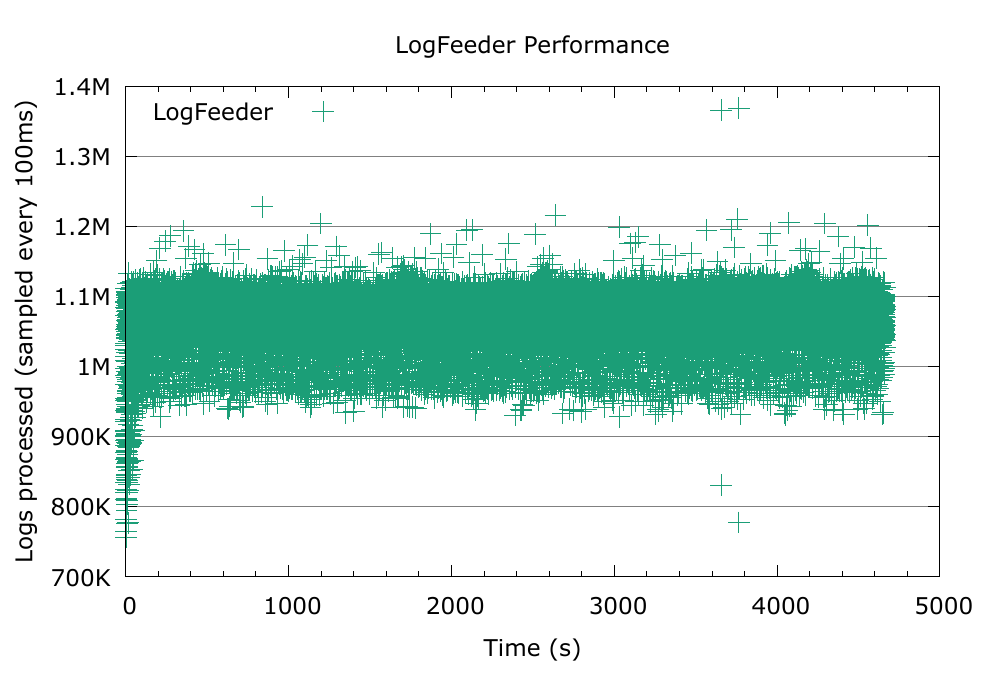}
\caption{LogFeeder performance using 4 header threads with logs of 219 Bytes}
\label{fig:log_feeder_performance}
\end{center}
\end{figure}

In our second test we wished to assess if the throughput was stable. To this end, we left {\em LogFeeder} running for more than an hour, measuring the number of logs passing through the system every 100 ms. The results are shown in Figure~\ref{fig:log_feeder_performance}, where we can see that the throughput remains remarkably stable with time.

\subsection{Log Database nodes}
\label{sec:storagenodes}

We tested a log database node using a server with two \texttt{Intel Xeon E5-2620 v3 @ 2.40 GHz} with 6 cores each. The server featured 64 GB of memory and several hard drives of 7200 RPM. The write processes were always ran with the \textit{ionice} command, specifying a real time priority, which is the maximum possible.

\begin{figure}[!b]
\begin{center}
\includegraphics[width=\columnwidth]{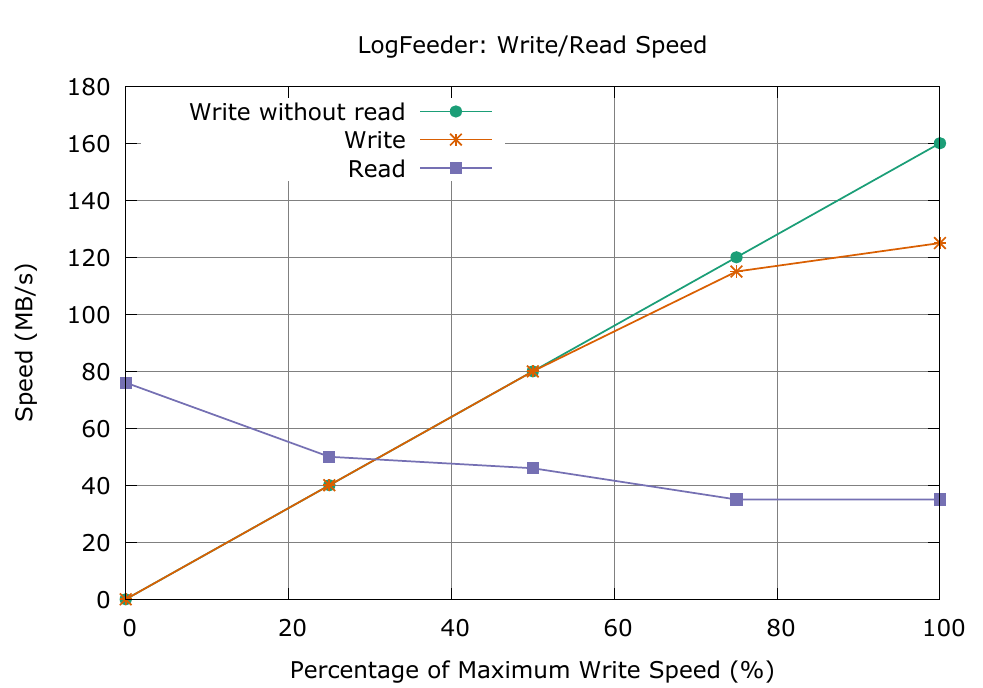}
\caption{LogFeeder Write/Read speed}
\label{fig:log_feeder_io_test}
\end{center}
\end{figure}

The first goal of this test was to check if the database process could take advantage of the full read/write capabilities of a hard drive. To this end, we first wrote 1 GB files directly from memory and determined that the maximum write speed of the drive was 170~MB/s. Then, we launched a database process receiving logs from a \textit{LogFeeder} and the resulting write speed was 160~MB/s (see Figure~\ref{fig:log_feeder_io_test}), mainly due to the additional pre-processing.

After that, we wished to find out the maximum write speed to be used so that read operations do not affect the performance. The priority of the read process was set to idle with the \textit{ionice} command. With this priority, I/O operations are only executed if there is no other process accessing the disk.

Consequently, we started the database process by limiting the write speed to 25\%, 50\%, 75\% and 100\% of the maximum write speed, which corresponds to 40 MB/s, 80 MB/s, 120 MB/s and 160 MB/s respectively. After the write process is stabilized, we executed a long read operation, while measuring the new write and read speeds. We also tested a use case reading from the database without a write process running at the same time. The results are shown in Figure~\ref{fig:log_feeder_io_test}.

We note that using more that 50\% (80~MB/s) of the maximum write speed reduces the performance of the write process when a long read is executed. Thanks to this experiments we can determine that the number of hard drives necessary for processing 3 million lines of log per second is equal to 7 when using logs of 117 bytes, and 14 drives for lengths of 291 bytes, plus the 64 bytes header used by our database. The lengths of 117 and 291 bytes are, respectively, the median and the 99th percentile of the log size found in Figure~\ref{fig:ecdf}.

\subsection{Pre-processing}
\label{sec:preprocesstest}
As explained in the previous section, the pre-processing scripts are completely malleable and hence, the performance of each script depends on the process to be performed on each type of log, call it filtering, aggregating and/or transforming the log lines.
Notwithstanding, we tried different ways of serializing log lines, (since it is the only mandatory pre-process before the logs are sent to {\em LoginsonReceptor}), with custom scripts, sending an Apache log file of 1 million lines, 100 bytes long each line, and measuring how long each program took to serialize the corpus. We chose different languages such as Flex, AWK and C, keeping in mind that we do not need them to process 3 million of log entries per second owing to the distributed processing of each kind of log with its corresponding AWK script.

We note that Flex and Bison, a lexical analyzer and parser generator respectively, were used to define the patterns of each log line. Bison defines the structure of the line to be parsed and Flex the regular expressions of each field. These tools are often used to define custom semantics and develop compilers using patterns and regular expressions. Their major asset is that a pre-knowledge of the log format is defined with rules, patterns and sub-patterns with different priorities, supporting optional fields in a smarter manner. Then, the code can be compiled with g++. The resulting parsers are strong but complex due to the pattern-matching performed with the rules. We note that using these tools usually requires training.

A much more straightforward script was done with AWK language, which is fairly easier to use than Flex and Bison. An AWK program is composed of a sequence of pattern-action statements. Each log message is broken up into a series of fields using the given delimiters. AWK is also able to match regular expressions against strings. This flexibility and ease-of-use are good assets that make AWK a good candidate as the code is much simpler to understand and modify. The test script in the experiment simply splits the log with the provided separator and prints the line on JSON format, adding the field names. The code is interpreted without need of compiling. \textit{AWK} also has other high-level capabilities that make it ideal for the rest of pre-processing operations like transformations, filters, or aggregations thanks to its associative arrays and regular expression features.

Finally, we used C as our third option, with a program in which fields were assigned to a structure and then printed in JSON format, adding the field names too. Although the code does the same than the AWK script, the complexity is higher and the code needs to be compiled as in the Flex option. It also requires a more elaborate programming. 

\begin{table}[!bth]
\begin{center}
 \caption {Serialization Benchmarks}
 \label{tab:serialization_tests}
 \begin{tabular}{| P{3cm} | c | c |}
 \hline
 \thead{System} & \thead{Time (seconds)} & \thead{Lines per second} \\ \hline
 Flex    & 13.9 & 71,942  \\ \hline
 AWK    & 3.5 & 285,714  \\ \hline
 C    & 1.27 & 787,401  \\ \hline
 \end{tabular}
\end{center}
\end{table}
For the latter two programs, a further exercise of coding is necessary in order to consider interdependent and/or optional fields in the log message (if any) but, despite of this, the code is easier to handle for the system programmer, compared to using the Flex and Bison syntaxes.

As shown in the Table~\ref{tab:serialization_tests}, the three of them were faster than Logstash~(29.5K logs/s/thread) and FluentD~(13K log/s/thread) (also Figure~\ref{fig:logstash_fluentd_benchmark}) at the time of converting logs to JSON. From among the three options tested, Flex is the slowest parser due to the complexity of the rules used to perform pattern-matching. The C ad-hoc parser is more than 10 times faster than the first parser, but in spite of this, the \textit{AWK} program represents a good balance between flexibility and speed, being 4 times quicker than the lexical parser but also much easier to learn than its counterparts. 
\subsubsection{LoginsonReceptor}
\label{subsec:loginson_receptor}
We tested {\em LoginsonReceptor} and \textit{Elasticsearch} on a machine with an \texttt{Intel i7 @ 2.4 Ghz, 16GB of RAM} and a PCI-e SSD drive. A grand total of 30 million serialized lines were sent to {\em LoginsonReceptor} on a single \textit{Elasticsearch node} that were indexed at an average speed of 83K lines/s, with Elasticsearch as the bottleneck of the test. Being this an ingest rate 30x lower than the initial 3M log lines received by the system, we consider that the attained throughput is sufficient for a summarized real time statistic which can be used to detect errors, anomalies, or incidents on the system with the relevant messages extracted in previous stages.
\begin{figure*}[!bth]
\begin{center}
\includegraphics[width=0.9\columnwidth]{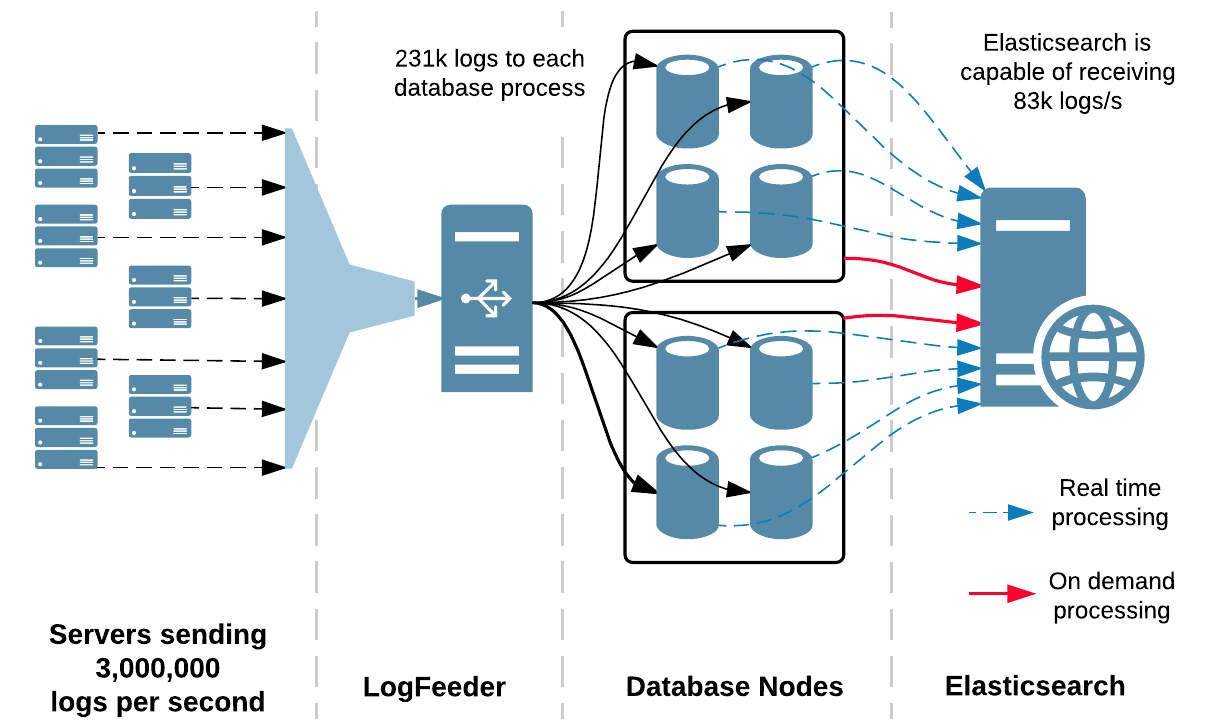}
\caption{Complete System}
\label{fig:completesystem}
\end{center}
\end{figure*}

\subsection{Complete system}
\label{subsec:complete_sys}

We tested the whole system in four servers. \textit{LogFeeder} was run in the same computer as the standalone test in section~\ref{sec:logfeeder}, concurrently with 13 database processes, matching the number of hard drives, were executed in two servers like the one used to test the storage speed in section~\ref{sec:storagenodes}. A fourth server was used to host \textit{Elasticsearch} and \textit{DBController}.

To match our requirements, \textit{LogFeeder} received 3 million log messages per second and splitted them between the database processes, that were writing to the disks at 50\% of their capacity. While \textit{LogFeeder} was inserting logs in the database nodes, we started a long read of the previously stored data. With this test we confirmed that the read operations did not interfere with the write operations. The logs were processed and received correctly in \textit{ElasticSearch}, where they could be visualized in \textit{Kibana}. 
Figure~\ref{fig:completesystem} shows how the test was executed.

In the last test we executed \textit{LogFeeder} and a database node in the same server used for the standalone test in Section~\ref{sec:logfeeder}. Such server has a 10~disk RAID~0 capable of storing data at 10~Gbps. \textbf{We assessed that the {\em Loginson} system was capable of receiving logs, storing them in the RAID and processing them at network rate (also 10~Gbps)}. However, in this test the write speed was affected when reading from the RAID, as the upper limit of 50\% of the write throughput of the RAID was exceeded, as stated in section~\ref{sec:storagenodes}.

\section{Conclusions}
\label{sec:conclusions}
In this paper we presented a high-performance solution for log collection, storage, processing and visualization called {\em Loginson}, which tradesoff simplicity with throughput, achieving a remarkable throughput of 3 million logs per second for the most common log size, and up to 10 million logs per second for small logs. Furthermore, all components in the solution are independent, which makes it possible to scale up horizontally for even higher speeds. 

As future work, we plan to have a better pre-processing system, as the current one is only capable of transforming the data for \textit{Elasticsearch} consumption. Such system would be capable of analyzing the log data to extract only the relevant information and detect anomalies in a timely manner.

\balance

\section{Acknowledgments}
\label{sec:ack}
This work has been partially supported by the \textit{Spanish Ministry of Economy and Competitiveness and the European Regional Development Fund} under the project \textbf{TR\'AFICA (MINECO/FEDER TEC2015-69417-C2-1-R)}. 

We would also like to thank Paloma Dominguez-Tom\'as for the effort of designing a logo for {\em Loginson}.

\nocite{*}

\bibliographystyle{spbasic}      


\end{document}